\documentclass{article}

\usepackage{times}
\usepackage{graphicx} 
\usepackage{subfigure} 

\usepackage{amsmath,amssymb,amsthm,bbm}

\usepackage{mathtools}
\newcommand{\myeq}{\stackrel{\mathclap{\scriptsize\mbox{\normalfont def.}}}{=}} 

\newtheorem{theorem}{Theorem}[section]

\newtheorem{proposition}{Proposition}[section]
\newtheorem{lemma}[theorem]{Lemma}

\theoremstyle{definition}
\newtheorem{definition}{Definition}[section]
 
\theoremstyle{remark}
\newtheorem{remark}{Remark}

\usepackage{natbib}

\usepackage{algorithm}
\usepackage{algorithmic}

\usepackage{hyperref}

\usepackage[accepted]{icml2017}

\icmltitlerunning{A Simulated Annealing Based Inexact Oracle for Wasserstein Loss Minimization}

\begin{document} 


\twocolumn[
\icmltitle{A Simulated Annealing Based Inexact Oracle \break for Wasserstein Loss Minimization}



\icmlsetsymbol{equal}{*}

\begin{icmlauthorlist}
\icmlauthor{Jianbo Ye}{psuist}
\icmlauthor{James Z. Wang}{psuist}
\icmlauthor{Jia Li}{psustat}
\end{icmlauthorlist}

\icmlaffiliation{psuist}{College of Information Sciences and Technology, 
The Pennsylvania State University, University Park, PA.}
\icmlaffiliation{psustat}{Department of Statistics, The Pennsylvania State University, University Park, PA.}

\icmlcorrespondingauthor{Jianbo Ye}{jxy198@ist.psu.edu}

\icmlkeywords{Wasserstein Distance, Simulated Annealing, Gibbs Sampling, Inexact Oracle}

\vskip 0.3in
]



\printAffiliationsAndNotice{}  

\begin{abstract} 
Learning under a Wasserstein loss, a.k.a. Wasserstein loss minimization (WLM), is an emerging research topic for gaining insights from a large set of structured objects. Despite being conceptually simple, WLM problems are computationally challenging because they involve minimizing over functions of quantities ({\it i.e.} Wasserstein distances) that themselves require numerical algorithms to compute. In this paper, we introduce a stochastic approach based on simulated annealing for solving WLMs. Particularly, we have developed a Gibbs sampler to approximate effectively and efficiently the partial gradients of a sequence of Wasserstein losses. Our new approach has the advantages of numerical stability and readiness for warm starts. These characteristics are valuable for WLM problems that often require multiple levels of iterations in which the oracle for computing the value and gradient of a loss function is embedded. We applied the method to optimal transport with Coulomb cost and the Wasserstein non-negative matrix factorization problem, and made comparisons with the existing method of entropy regularization. 
\end{abstract} 

\section{Introduction}
An oracle is a computational module in an optimization procedure that is applied iteratively to obtain certain characteristics of the function being optimized. Typically, it
calculates the value and gradient of loss function $l(\mathbf x,\mathbf y)$. 
In the vast majority of machine learning models, where those loss functions 
are decomposable along each dimension ({\it e.g.}, $L_p$ norm, KL divergence, or hinge loss), 
$\nabla_{\mathbf x}{l}(\cdot, \mathbf y)$ or $\nabla_{\mathbf y}{l}(\mathbf x,\cdot)$ is computed 
in $O(m)$ time, $m$ being the complexity of outcome variables $\mathbf x$ or $\mathbf y$. 
This part of calculation is often negligible compared with the calculation of 
full gradient with respect to the model parameters.
But this is no longer the case in learning problems based on Wasserstein distance due to the intrinsic complexity of the distance.
We will call such problems {\em Wasserstein loss minimization} (WLM). 
Examples of WLMs include Wasserstein barycenters~\cite{li2008real,agueh2011barycenters,cuturi2014fast,benamou2015iterative,ye2014scaling,ye2015accelerated}, 
principal geodesics~\cite{seguy2015principal}, 
nonnegative matrix factorization~\cite{roletfast,sandler2009nonnegative}, 
barycentric coordinate~\cite{bonneel2016wasserstein},
and multi-label classification~\cite{frogner2015learning}.

{Wasserstein distance} is defined as the cost of matching
two probability measures, originated from the literature of optimal transport (OT)~\cite{monge1781memoire}.
It takes into account the cross-term similarity between different support points of the distributions, a level of complexity beyond the usual vector data treatment, {\it i.e.}, to convert the distribution into a vector of frequencies. It has been promoted for comparing sets of vectors ({\it e.g.} bag-of-words models) by researchers in computer vision, multimedia
and more recently natural language processing~\cite{kusner2015word,ye2017determining}. 
However, its potential as a powerful loss function for machine learning has been underexplored. The major obstacle is a lack of standardized and robust numerical methods to solve WLMs. Even to empirically better understand the advantages of the distance is of interest.

As a long-standing consensus, solving WLMs is 
challenging~\cite{cuturi2014fast}. 
Unlike the usual optimization in machine learning where the loss and the (partial) gradient
can be calculated in linear time, these quantities are non-smooth and hard to obtain in WLMs,
requiring solution of a costly network transportation problem (a.k.a. OT).
The time complexity, $O(m^3\log m)$, is prohibitively high~\cite{orlin1993faster}. In contrast to the $L_p$ or KL counterparts, 
this step of calculation elevates from a negligible fraction of the overall learning problem to a dominant portion, preventing the scaling of WLMs to large data. 
Recently, iterative approximation techniques have been developed 
to compute the loss and the (partial) gradient at 
complexity $O(m^2/\varepsilon)$~\cite{cuturi2013sinkhorn,wang2014bregman}. 
However, nontrivial algorithmic efforts are needed to incorporate
these methods into WLMs because WLMs often require multi-level loops~\cite{cuturi2014fast,frogner2015learning}.
Specifically, one must re-calculate through many iterations the loss and its partial gradient in order
to update other model dependent parameters.

We are thus motivated to seek for a fast {\it inexact} oracle that (i) runs at lower time complexity per iteration, and
(ii) accommodates warm starts and meaningful early stops. These two properties are equally important for 
efficiently obtaining adequate approximation to the solutions of a sequence of slowly changing OTs. 
The second property ensures that the subsequent OTs can effectively leverage the solutions of the earlier OTs 
so that the total computational time is low. 
Approximation techniques with low complexity per iteration
already exist for solving a single OT, but they do not possess the second property. In this paper, we introduce a method that uses a time-inhomogeneous Gibbs sampler as an inexact oracle for Wasserstein losses. The Markov chain Monte Carlo (MCMC) based method naturally satisfies the second property, as reflected by the intuition of physicists that MCMC samples can efficiently ``remix from a previous equilibrium.'' 

We propose a new optimization approach
based on {Simulated Annealing} (SA)~\cite{kirkpatrick1983optimization,corana1987minimizing}
for WLMs where the outcome variables are treated as probability measures. 
SA is especially suitable for the dual OT problem, where the usual Metropolis sampler
can be simplified to a Gibbs sampler. To our knowledge, existing optimization techniques used on WLMs are different from MCMC. In practice, MCMC is known to easily accommodate warm start, which is particularly useful in the context of WLMs. We name this approach \textit{Gibbs-OT} for short. 
The algorithm of Gibbs-OT is as simple and efficient as the Sinkhorn's algorithm
--- a widely accepted method to approximately solve OT~\cite{cuturi2013sinkhorn}.  
We show that Gibbs-OT enjoys improved numerical stability and several algorithmic characteristics valuable for general WLMs. 
By experiments, we demonstrate the effectiveness of Gibbs-OT for solving optimal transport with Coulomb cost~\cite{benamou2016numerical} and 
the Wasserstein non-negative matrix factorization (NMF) problem~\cite{sandler2009nonnegative,roletfast}. 

\section{Related Work}
Recently, several methods have been proposed  
to overcome the aforementioned difficulties in solving WLMs. Representatives include entropic regularization~\cite{cuturi2013sinkhorn,cuturi2014fast,benamou2015iterative} and Bregman ADMM~\cite{wang2014bregman,ye2015accelerated}. 
The main idea is to augment the original optimization problem with a strongly convex term such that the regularized objective becomes a smooth function of all its coordinating parameters. 
Neither the Sinkhorn's algorithm nor Bregman ADMM can be readily integrated into a general WLM. 
Based on the entropic regularization of primal OT, 
\citet{cuturi2015smoothed} recently showed that the Legendre transform of the entropy regularized 
Wasserstein loss and its gradient can be computed in closed form, which appear in the first-order condition of some complex WLM problems. Using this technique, the regularized primal problem can be converted to an equivalent Fenchel-type dual problem that has a faster numerical solver in the Euclidean space~\cite{roletfast}. 
But this methodology can only be applied to a certain class of WLM problems of which the Fenchel-type dual has closed forms of objective and full gradient.
In contrast, the proposed SA-based approach directly deals with the dual OT problem without assuming any particular mathematical structure of the WLM problem, and hence is more flexible to apply. 

More recent approaches base on solving the dual OT problems have been proposed 
to calculate and optimize the Wasserstein distance between a {\em single pair}
of distributions with very large support sets --- 
often as large as the size of an entire machine learning dataset~\cite{montavon2016wasserstein,genevay2016stochastic,arjovsky2017wasserstein}. 
For these methods, scalability is achieved in terms of the support size.
Our proposed method has a different focus on
calculating and optimizing Wasserstein distances between {\em many pairs} all together in WLMs, with
each distribution having a moderate support size ({\it e.g.}, dozens or hundreds).
We aim at scalability for the scenarios when a large set of distributions have to be handled simultaneously, that is, the optimization cannot be decoupled on the distributions. In addition, existing methods have no on-the-fly mechanism to control the approximation quality at a limited number of iterations.

\section{Preliminaries of Optimal Transport}
In this section, we present notations, mathematical backgrounds, and set up the problem of interest. 
\begin{definition}[Optimal Transportation, OT]
\label{def:wass}
Let $\mathbf p\in \Delta_{m_1}, \mathbf q\in \Delta_{m_2}$,
where $\Delta_m$ is the set of $m$-dimensional simplex:
$\Delta_m \myeq \{\mathbf q \in \mathbb R_+^{m}: \langle \mathbf q, \mathbbm{1}\rangle = 1\}$.
The set of transportation plans between
$\mathbf p$ and $\mathbf q$ is defined as $\Pi(\mathbf p,\mathbf q)\myeq 
\{Z\in\mathbb R^{m_1\times m_2}: Z \cdot \mathbbm{1}_{m_2}= \mathbf p; 
Z^T \cdot \mathbbm{1}_{m_1}= \mathbf q;\}$.
Let $M\in \mathbb R_+^{m_1 \times m_2}$ be the matrix of costs. 
The optimal transport cost between $\mathbf p$ and $\mathbf q$ with respect to $M$ is
\begin{equation}
W(\mathbf p,\mathbf q) \myeq \min_{Z\in \Pi(\mathbf p, \mathbf q)} \langle Z, M\rangle\;.\label{eq:lpprimal}
\end{equation}
In particular, $\Pi(\cdot, \cdot)$ is often called the coupling set. 
\end{definition}

Now we relate primal version of (discrete) OT to a variant of its dual version. One
may refer to~\citet{villani2003topics} for the general background of the 
Kantorovich-Rubenstein duality.  
In particular, our formulation introduces an auxiliary parameter $C_M$
for the sake of mathematical soundness in defining Boltzmann distributions. 
\begin{definition}[Dual Formulation of OT]
Let $C_M>0$, denote vector $[g_1,\ldots,g_{m_1}]^T$ by $\mathbf g$, and vector $[h_1,\ldots,h_{m_2}]^T$ by $\mathbf h$. 
We define the dual domain of OT by
\begin{multline}
\Omega (M) \myeq\Big\{\left.\mathbf f=[\mathbf g; \mathbf h]\in \mathbb R^{m_1+m_2}~\right|~\\
-C_{M}<g_i - h_j \le M_{i,j}, 
1\le i\le m_1, 1\le j \le m_2\Big\}\;.\label{eq:omega}
\end{multline}
Informally, for a sufficiently large $C_M$ (subject to $\mathbf p,\mathbf q, M$), 
the LP problem Eq.~\eqref{eq:lpprimal} can be reformulated as
\footnote{However, for any proper $M$ and strictly positive $\mathbf p, \mathbf q$, 
there exists $C_M$ such that the optimal value of primal problem 
is equal to the optimal value of the dual problem.
This modification is solely for an ad-hoc treatment of a single OT problem.
In general cases of $(\mathbf p, \mathbf q, M)$, 
when $C_M$ is pre-fixed, the solution of Eq.~\eqref{eq:dual} may be suboptimal. }
\begin{equation}
W(\mathbf p, \mathbf q)=\max_{\mathbf f\in \Omega(M)}
\langle \mathbf p, \mathbf g\rangle - \langle \mathbf q, \mathbf h\rangle\;.\label{eq:dual}
\end{equation}
\end{definition}
Let the optimum set be $\Omega^\ast(M)$. Then any optimal point $\mathbf f^\ast=(\mathbf g^\ast, \mathbf h^\ast)\in 
\Omega^\ast(M)$ constructs a (projected) subgradient such that $\mathbf g^\ast\in \partial W/\partial \mathbf p $
and $-\mathbf h^\ast \in \partial W / \partial \mathbf q $~. The main computational difficulty
of WLMs comes from the fact that (projected) subgradient $\mathbf f^\ast$ is not efficiently solvable. 

Note that $\Omega(M)$ is an unbound set in $\mathbb R^{m_1+m_2}$. In order to constrain the feasible region
to be bounded, we alternatively define 
\begin{equation}
\Omega_0(M)\!=\!\{\mathbf f=[\mathbf g; \mathbf h]\in \Omega(M) ~|~ g_1 = 0\}.
\end{equation} 
One can show that the maximization in $\Omega(M)$ as Eq.~\eqref{eq:dual} is 
equivalent to the maximization in $\Omega_0(M)$ because 
$\langle\mathbf p, \mathbbm 1_{m_1}\rangle = \langle \mathbf q, \mathbbm 1_{m_2}\rangle$. 

\section{Simulated Annealing for Optimal Transport via Gibbs Sampling}\label{sec:gibbs}
Following the basic strategy outlined in the seminal paper of simulated annealing~\citep{kirkpatrick1983optimization},
we present the definition of Boltzmann distribution supported on $\Omega_0(M)$ 
below which, as we will elaborate, links the dual formulation of OT to a Gibbs sampling scheme (Algorithm~\ref{alg:sa-ot} below). 

\begin{definition}[Boltzmann Distribution of OT]\label{def:boltzmann}
Given a temperature parameter $T>0$, the Boltzmann distribution of OT
is a probability measure on $\Omega_0(M)\subseteq \mathbb R^{m_1+m_2-1}$ such that 
\begin{equation}
p(\mathbf f; \mathbf p, \mathbf q) \propto 
\exp \left[\frac{1}{T} \left(\langle \mathbf p, \mathbf g\rangle - \langle \mathbf q, \mathbf h\rangle \right)\right]. \label{eq:pdf}
\end{equation}
It is a well-defined probability measure for an arbitrary finite $C_M>0$.
\end{definition}

The basic concept behind SA states that
the samples from the Boltzmann distribution will eventually concentrate at the optimum set
of its deriving problem ({\it e.g.} $W(\mathbf p,\mathbf q)$) as $T\rightarrow 0$.
However, since the Boltzmann distribution is often difficult to sample, 
a practical convergence rate remains mostly unsettled for specific MCMC methods.

Because $\Omega (M)$ defined by Eq.~\eqref{eq:omega} (also $\Omega_0$)
has a conditional independence structure among variables, a Gibbs sampler can be naturally applied to the Boltzmann 
distribution defined by Eq.~\eqref{eq:pdf}.
We summarize this result below.
\begin{proposition}
Given any $\mathbf f=(\mathbf g; \mathbf h)\in \Omega_0(M)$ and any $C_M>0$, we have for any $i$ and $j$,
\begin{eqnarray}
g_i \le U_i(\mathbf h) & \myeq & \min_{1\le j\le m_2} \left( M_{i,j} + h_j\right), \label{eq:bounds1} \\
h_j \ge L_j(\mathbf g) & \myeq & \max_{1\le i\le m_1} \left( g_i - M_{i,j}\right). \label{eq:bounds2}
\end{eqnarray}
and 
\begin{eqnarray}
g_i > \widehat{L}_i(\mathbf h) & \myeq & \max_{1\le j\le m_2} \left( -C_M + h_j\right), \label{eq:bounds3} \\
h_j < \widehat{U}_j(\mathbf g) & \myeq & \max_{1\le i\le m_1} \left( C_M + g_i\right). \label{eq:bounds4}
\end{eqnarray}
Here $U_i=U_i(\mathbf h) $ and $L_j=L_j(\mathbf g)$ are auxiliary variables.
Suppose $\mathbf f$ follows the Boltzmann distribution by Eq.~\eqref{eq:pdf},
$g_i$'s are conditionally independent given $\mathbf h$, and likewise
$h_j$'s are also conditionally independent given $\mathbf g$. Furthermore, it is immediate from
Eq.~\eqref{eq:pdf} that each of their conditional probabilities within its feasible region (subject to $C_M$) 
satisfies 
\begin{flalign}
&p(g_i | \mathbf h) \propto \exp\left(\dfrac{g_i p_i}{T}\right), 
\; \widehat{L}_i(\mathbf h) < g_i \le U_i(\mathbf h),\\ 
&p(h_j | \mathbf g) \propto \exp\left(-\dfrac{h_j q_j}{T}\right), 
\; {L}_j(\mathbf g) \le h_j < \widehat{U}_j(\mathbf g),
\end{flalign}\label{prop:cond}
where $2\le i\le m_1$ and $1\le j\le m_2$. 
\end{proposition}
\begin{remark}
As $C_M\rightarrow +\infty$, 
$\widehat{U}_j(\mathbf g)\rightarrow +\infty$ 
and $\widehat{L}_i(\mathbf h)\rightarrow -\infty$. 
For $2 \le i \le m_1$ and $1 \le j \le m_2$,
one can approximate the conditional probability 
$p(g_i | \mathbf h)$ and $p(h_j | \mathbf g)$ by exponential distributions. 
\end{remark}
By Proposition.~\ref{prop:cond},  
our proposed time-inhomogeneous Gibbs sampler is given in Algorithm~\ref{alg:sa-ot}.
\begin{algorithm}[t]
Given $\mathbf f^{(0)}\in \Omega_0(M)$, $\mathbf p\in \Delta_{m_1}$ and $\mathbf q\in \Delta_{m_2}$, 
and $T^{(1)},\ldots,T^{(2N)}> 0$, for $t=1,\ldots,N$, we define the following Markov chain
\begin{enumerate}
\item Randomly sample 
\[\theta_1,\ldots,\theta_{m_2}\, \overset{i.i.d.}{\sim}\, \mbox{Exponential}(1).\]
For $j=1,2,\ldots,m_2$, let
\begin{equation}
\begin{cases}
L_j^{(t)} := & \max_{1\le i\le m_1} \left( g_i^{(t-1)} - M_{i,j}\right)\\
h_j^{(t)} :=& L_{j}^{(t)} + \theta_j \cdot T^{(2t-1)} / q_j
\end{cases}\label{eq:update1}
\end{equation}
\item Randomly sample 
\[\theta_2,\ldots,\theta_{m_1}\, \overset{i.i.d.}{\sim}\, \mbox{Exponential}(1).\]
For $i=(1), 2,\ldots,m_1$, let
\begin{equation}
\begin{cases}
U_i^{(t)} := & \min_{1\le j\le m_2} \left( M_{i,j} + h_j^{(t)}\right) \\
g_i^{(t)} := & U_{i}^{(t)} - \theta_i \cdot T^{(2t)} / p_i
\end{cases}\label{eq:update2}
\end{equation}
\end{enumerate}
\caption{Gibbs Sampling for Optimal Transport}\label{alg:sa-ot}
\vspace{-1em}
\end{algorithm}
Specifically in Algorithm~\ref{alg:sa-ot}, the variable $g_1$ is fixed to zero by the definition
of $\Omega_0(M)$.  But we have found in experiments that by calculating $U_1^{(t)}$ and sampling $g_1^{(t)}$
in Algorithm~\ref{alg:sa-ot} according to Eq.~\eqref{eq:update2}, one can still generate MCMC
samples from $\Omega(M)$ such that the energy quantity $\langle \mathbf p, \mathbf g\rangle - \langle \mathbf q, \mathbf h\rangle$ converges to 
the same distribution as that of MCMC samples from $\Omega_0(M)$. 
Therefore, we will not assume $g_1=0$ from now on and develop analysis solely for the unconstrained version of Gibbs-OT. 

\begin{figure*}[htp]\vskip 0.2in
\centering
\subfigure[20 iterations]{\includegraphics[width=0.32\textwidth, trim=2cm 7cm 2cm 7cm, clip]{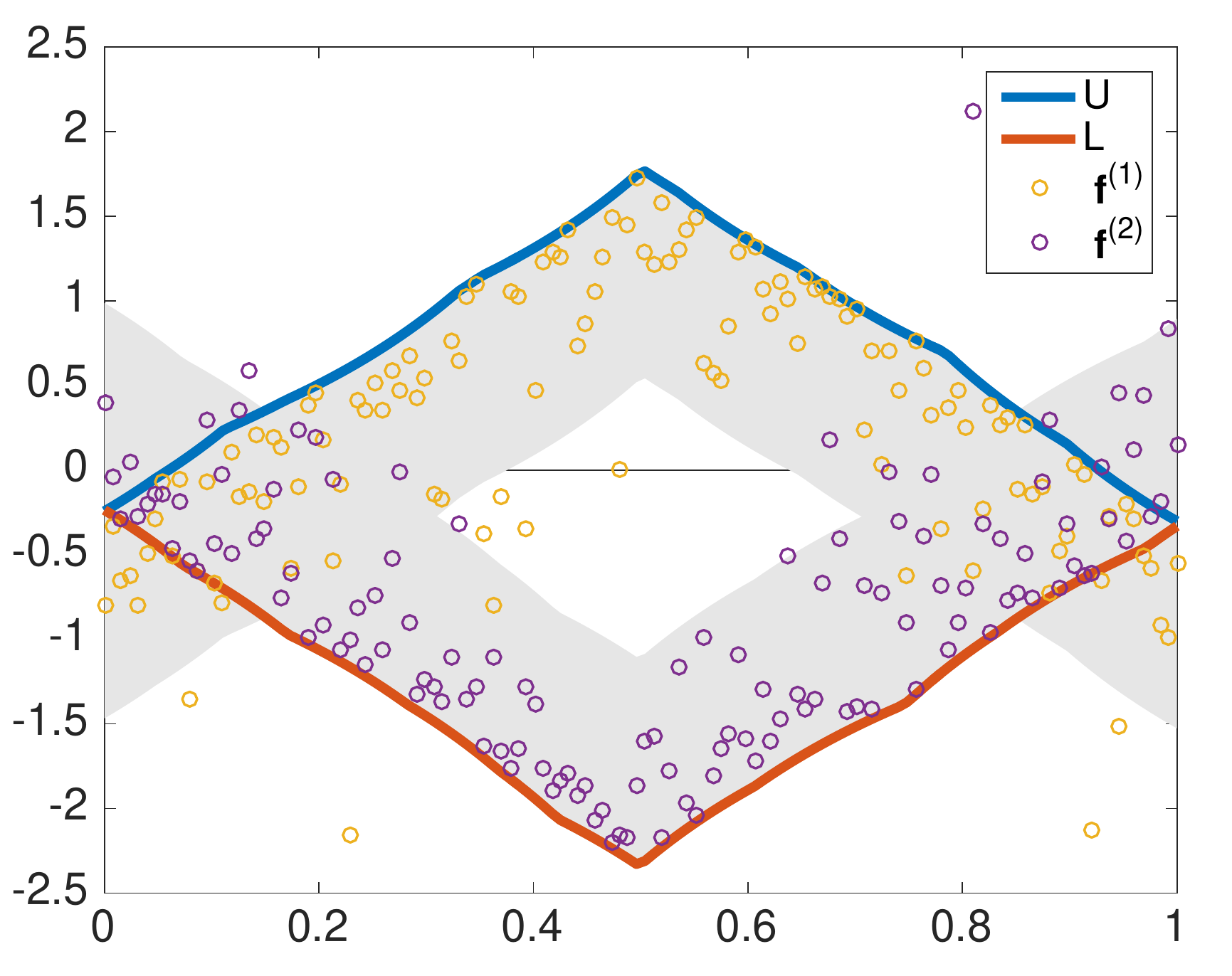}}
\subfigure[40 iterations]{\includegraphics[width=0.32\textwidth, trim=2cm 7cm 2cm 7cm, clip]{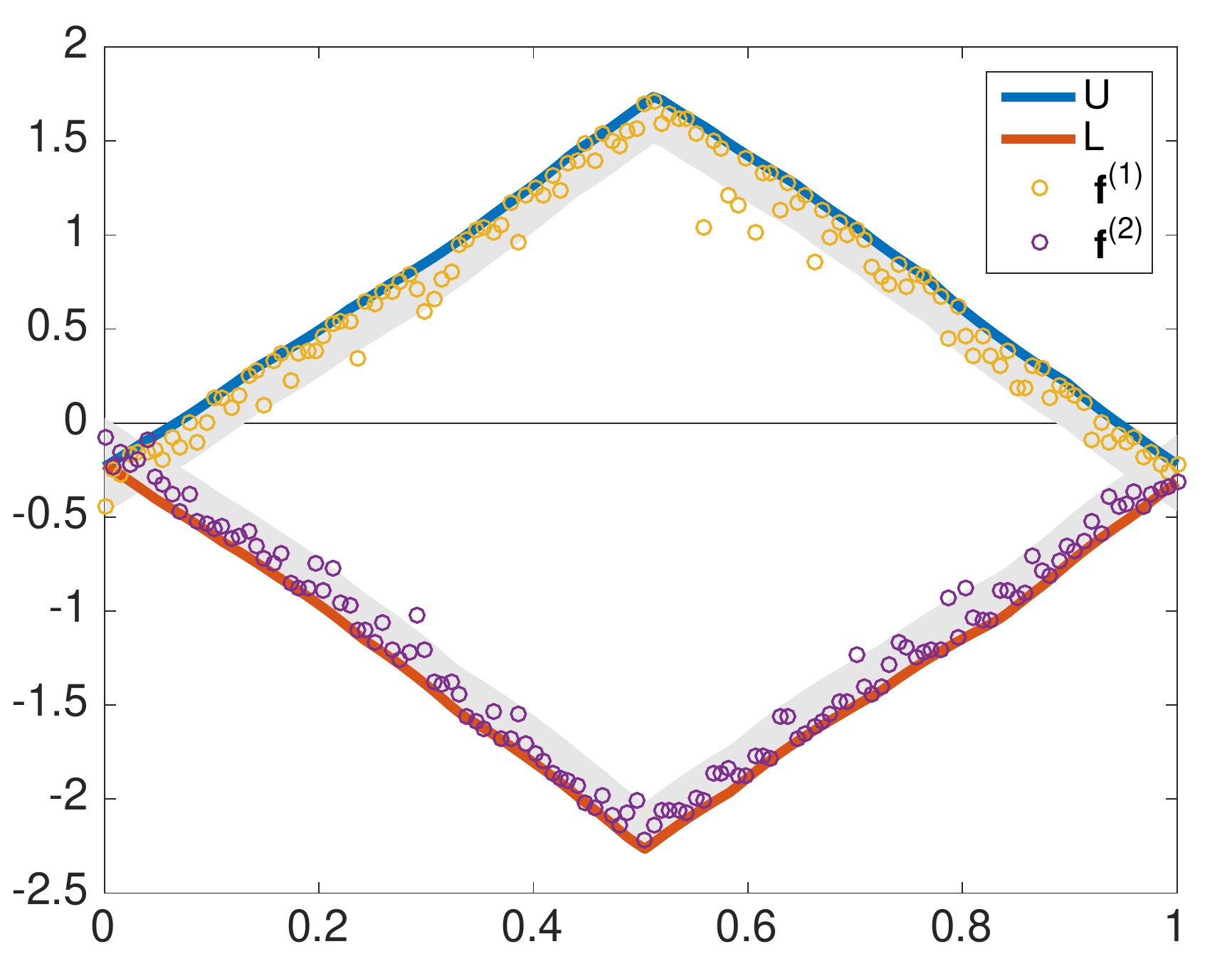}}
\subfigure[60 iterations]{\includegraphics[width=0.32\textwidth, trim=2cm 7cm 2cm 7cm, clip]{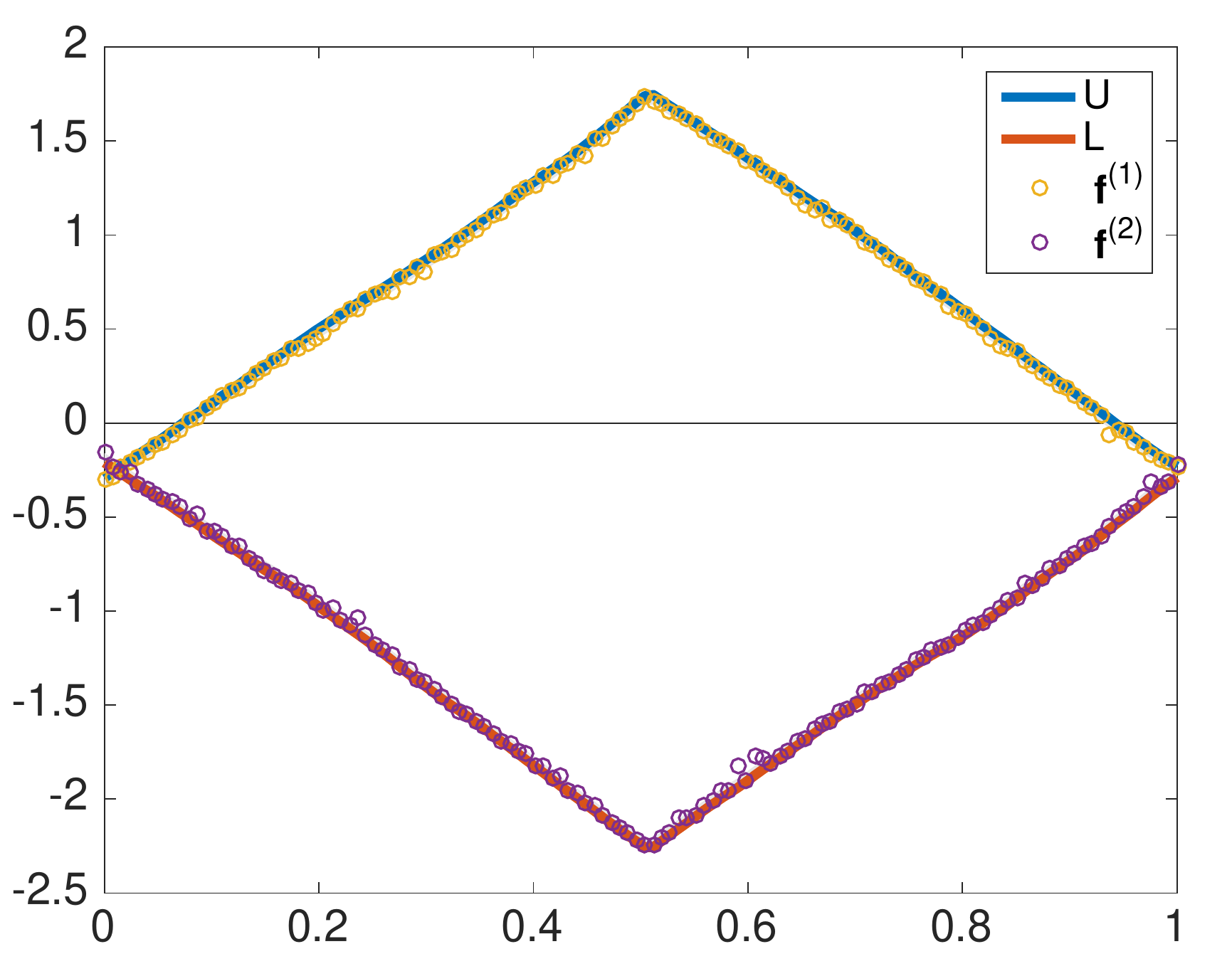}}
\caption{The Gibbs sampling of the proposed SA method. From left to right is
an illustrative example of a simple 1D optimal transportation problem with Coulomb cost and plots
of variables for solving this problem at different number of iterations 
$\in \{20, 40, 60\}$ using the inhomogeneous Gibbs sampler. Particularly, the 95\% percentile of the 
exponential distributions are marked by the gray area.}\label{fig:sa}
\vskip -0.2in
\end{figure*}

Figure~\ref{fig:sa} illustrates the behavior of the proposed Gibbs sampler with a cooling schedule at different temperatures. As $T$ decreases along iterations, the 95\% percentile band for sample $\mathbf f$ becomes thinner and thinner. 

\begin{remark}
Algorithm~\ref{alg:sa-ot} does not specify the actual cooling schedule, nor does the analysis
of the proposed Gibbs sampler in Theorem~\ref{bounds}. We have been agnostic here for a reason.
In the SA literature, cooling schedules with guaranteed optimality are often too slow to be useful in practice. To our knowledge, the guaranteed rate of SA approach is worse than the combinatorial solver for OT. As a result, a well-accepted practice of SA for many complicated optimization problems is to empirically adjust cooling schedules, a strategy we take for our experiments. 
\end{remark}

\begin{remark}
Although the exact cooling schedule is not specified, we still provide a quantitative 
upper bound of the chosen temperature $T$ at different iterations in Appendix~\ref{sec:theory} Eq.~\eqref{eq:tbound}.
One can calculate such bound at the cost of $m\log m$ at certain iterations to check
whether the current temperature is too high for the used Gibbs-OT to
accurately approximate the Wasserstein gradient. 
In practice, we find this bound helps one quickly select
the beginning temperature of Gibbs-OT algorithm. 
\end{remark}

\begin{definition}[Notations for Auxiliary Statistics]
Besides the Gibbs coordinates $\mathbf g$ and $\mathbf h$, 
the Gibbs-OT sampler naturally introduces two auxiliary variables, $\mathbf U$ and $\mathbf L$. 
Let $\mathbf L^{(t)}=\left[L_1^{(t)},\ldots,L_{m_2}^{(t)}\right]^T$ and $\mathbf U^{(t)}=\left[U_1^{(t)},\ldots,U_{m_1}^{(t)}\right]^T$.
Likewise, denote the collection of $g_i^{(t)}$ and $h_j^{(t)}$ by vectors $\mathbf g^{(t)}$ and $\mathbf h^{(t)}$ respectively. 
The following sequence of auxiliary statistics 
\begin{multline}
[\ldots, \mathbf z^{2t-1}, \mathbf z^{2t},\mathbf z^{2t+1},\ldots,]\myeq \\
\left[\ldots,\begin{bmatrix}\mathbf L^{(t)}\\ \mathbf U^{(t-1)}\end{bmatrix},\begin{bmatrix}\mathbf L^{(t)}\\ \mathbf U^{(t)}\end{bmatrix},\begin{bmatrix}\mathbf L^{(t+1)}\\ \mathbf U^{(t)}\end{bmatrix},\ldots\right]\label{eq:z}
\end{multline}
for $t=1,\ldots,N$ is also a Markov chain. They can be redefined equivalently by specifying the transition probabilities $p(\mathbf z^{n+1}|\mathbf z^{n})$ for $n=1,\ldots,2N$, a.k.a., the conditional p.d.f. $p(\mathbf U^{(t)} | \mathbf L^{(t)})$ 
for $t=1,\ldots,N$ and $p(\mathbf L^{(t+1)} | \mathbf U^{(t)})$ for $t=1,\ldots,N-1$. 
\end{definition}

One may notice that the alternative representation converts the Gibbs sampler to one 
whose structure is similar to a hidden Markov model, where the $\mathbf g,\mathbf h$ chain
is conditional independent given the $\mathbf U,\mathbf L$ chain and has (factored) exponential emission distributions.
We will use this equivalent representation in Appendix~\ref{sec:theory} and develop analysis based on 
the $\mathbf U,\mathbf L$ chain accordingly. 

\begin{remark}
We now consider the function 
\[ V(\mathbf x,\mathbf y )\myeq
\langle \mathbf p, \mathbf x\rangle - \langle \mathbf q,\mathbf y\rangle\;,\]
and define a few additional notations. Let $V(\mathbf U^{t'}, \mathbf L^{t})$ be denoted by
$V(\mathbf z^{t+t'})$, where $t'=t \mbox{ or } t\!-\!1$.
If $\mathbf g,\mathbf h$ are independently resampled 
according to Eq.~\eqref{eq:update1} and~\eqref{eq:update2},
we will have the inequalities that 
\[\mathbb E\left[V(\mathbf g,\mathbf h) | \mathbf z^n \right]
\le V(\mathbf z^{n})\;.\] 
Both $V(\mathbf z)$ and $V(\mathbf g, \mathbf h)$ converges to the exact loss $W(\mathbf p,\mathbf q)$ 
at the equilibrium of Boltzmann distribution $p(\mathbf f; \mathbf p,\mathbf q)$ as $T\to 0$.~\footnote{The conditional quantity 
$V(\mathbf z^{n}) - V(\mathbf g,\mathbf h) | \mathbf z^n$
is the sum of two Gamma random variables: $\mbox{Gamma}(m_1, 1/T^{(2t)}) + \mbox{Gamma}(m_2, 1/T^{(2t'+1)})$ where $t'=t$ or $t'=t-1$. }
\end{remark}

\section{Gibbs-OT: An Inexact Oracle for WLMs} 
In this section, we introduce a non-standard SA approach for the general WLM problems. 
The main idea is to replace the standard Boltzmann energy with an asymptotic consistent upper bound, 
outlined in our previous section.  Let
\[\mathfrak{R}(\theta) :=
\sum\limits_{i=1}^{\left|\mathcal D\right|} W(\mathbf p_i(\theta), \mathbf q_i(\theta))\]
be our prototyped objective function, 
where $\mathcal D$ represents a dataset, $\mathbf p_i,\mathbf q_i$ are prototyped probability 
densities for representing the $i$-th instance. 
We now discuss how to solve $\min_{\theta\in\Theta} \mathfrak{R}(\theta)$.

To minimize the Wasserstein losses $W(\mathbf p, \mathbf q)$ approximately in such WLMs,
we propose to instead optimize its asymptotic consistent upper bound 
$\mathbb E[V(\mathbf z)]$ at equilibrium of Boltzmann distribution 
$p(\mathbf f; \mathbf p, \mathbf q)$ using its stochastic gradients: 
$\mathbf U \in \partial V(\mathbf z)/\partial \mathbf p$ and
$-\mathbf L\in \partial V(\mathbf z)/\partial \mathbf q$~.
Therefore, one can calculate the gradient approximately:
\[\nabla_{\theta}{\mathfrak{R}} \approx \sum\limits_{i=1}^{\left|\mathcal D\right|} \left[
 J_{\theta}({\mathbf p_i}(\theta)) \mathbf U_i - J_{\theta}({\mathbf q_i}(\theta)) \mathbf L_i\right]\]
where $J_{\theta}(\cdot)$ is the Jacobian, $\mathbf U_i$, $\mathbf L_i$ are computed from
Algorithm~\ref{alg:sa-ot} for the problem $W(\mathbf p_i, \mathbf q_i)$ respectively. 
Together with the iterative updates of model parameters $\theta$, 
one gradually anneals the temperature $T$. The equilibrium of 
$p(\mathbf f; \mathbf p, \mathbf q)$ becomes more and more concentrated. 
We assume the inexact oracle at a relatively higher temperature is 
adequate for early updates of the model parameters,
but sooner or later it becomes necessary to set $T$ smaller to better approximate the exact loss. 

It is well known that the variance of stochastic gradient usually affects the rate of convergence. The reason to replace $V(\mathbf g, \mathbf h)$ with $V(\mathbf z)$ as the inexact oracle (for some $T>0$) is motivated by the same intuition. 
The variances of MCMC samples $g_i^{(t)}, h_j^{(t)}$ of Algorithm~\ref{alg:sa-ot} can be very large
if $p_i/T$ and $q_j/T$ are small, making the embedded first-order method inaccurate unavoidably. But we find the variances of max/min statistics $U_i^{(t)}, L_j^{(t)}$ are much smaller. Fig.~\ref{fig:sa} shows an example. The bias introduced in the replacement is also well controlled by decreasing the temperature parameter $T$. 
For the sake of efficiency, we use a very simple convergence diagnostics in the practice of Gibbs-OT. 
We check the values of $V(\mathbf z^{(2t)})$ such that the Markov chain is roughly considered mixed if every $\tau$ iteration the quantity $V(\mathbf z^{(2t)})$ (almost) stops increasing ($\tau\!\!=\!\! 5$ by default),
say, for some $t$, 
\[V(\mathbf z^{(2t)}) - V(\mathbf z^{(2(t-\tau))}) < 0.01 \tau T \cdot V(\mathbf z^{(2t)}),\] 
we terminate the Gibbs iterations.

\begin{figure}[ht!]
\includegraphics[width=.5\textwidth,trim={0.5cm 2.4cm 0.5cm 2.5cm},clip]{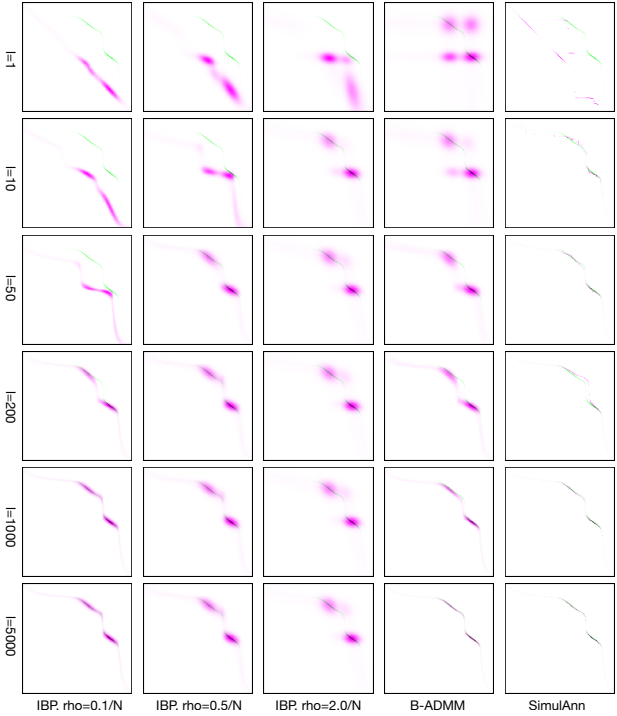}
\caption{A simple example for OT between two 1D distribution:
The solutions by Iterative Bregman Projection, B-ADMM, and Gibbs-OT are shown in pink, 
while the exact solution by linear programming is shown in green.
Images in the rows from top to bottom present results at different iterations 
$\{1, 10, 50, 200, 1000, 5000\}$;
The left three columns are by IBP with $\varepsilon=\{0.1/N, 0.5/N, 2/N\}$,
where $[0,1]$ is discretized with $N=128$ uniformly spaced points. 
The fourth column is by B-ADMM (with default parameter $\tau_0=2.0$).
The last column is the proposed Gibbs-OT, with a geometric cooling schedule.
With a properly selected cooling schedule, one can achieve 
fast convergence of OT solution without comprising much solution quality.}\label{fig:toy}
\end{figure}

\begin{figure*}[ht!]\centering
\includegraphics[width=\textwidth, trim=.5cm 11.2cm .5cm 10cm, clip]{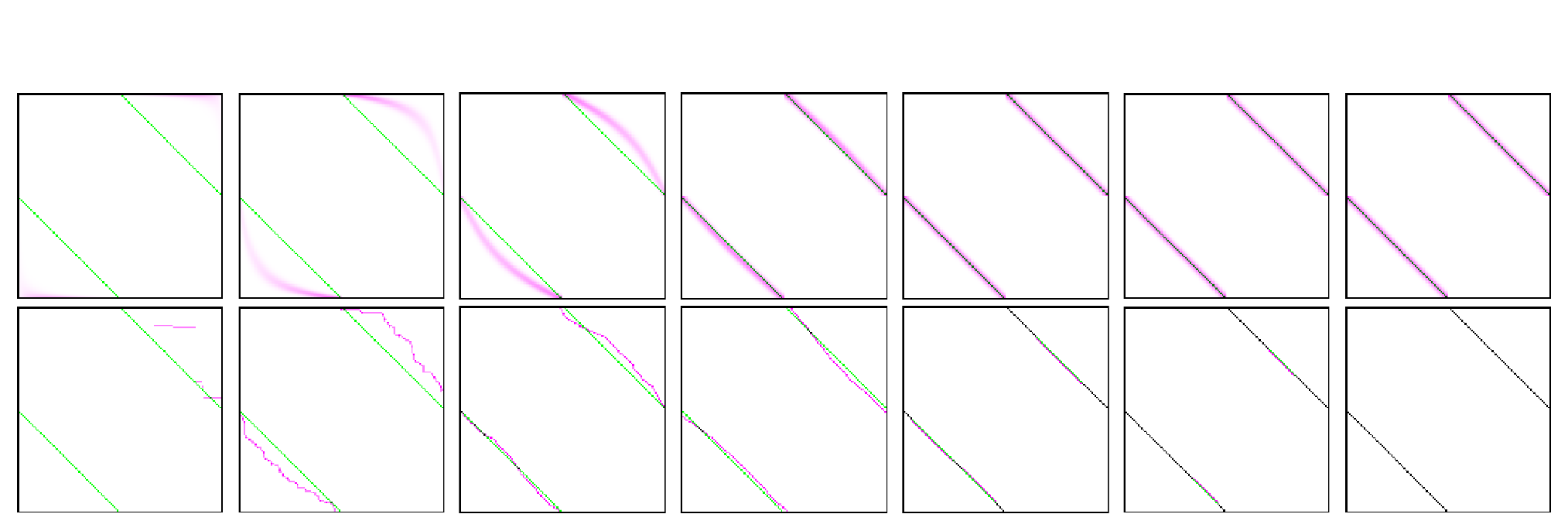}
\caption{The recovered primal solutions for two uniform 1D distribution with Coulumb cost.
The approximate solutions are shown in pink,
while the exact solution by linear programming is shown in green.
Top row: entropic regularization with $\varepsilon = 0.5/N$. Bottom row: Gibbs-OT. 
Images in the rows from left to right present results at different max iterations 
$\{1, 10, 50, 200, 1000, 2000, 5000\}$.}\label{fig:coulomb}
\end{figure*}

\section{Applications of Gibbs-OT}
\subsection{Toy OT Examples}

\textbf{1D Case with Euclidean Cost.} 
We first illustrate the differences between the approximate primal solutions computed by different methods
by replicating a toy example in~\cite{benamou2015iterative}. The toy example calculates the OT between
two 1D two-mode distributions. We visualize their solved coupling as a 2D image in Fig.~\ref{fig:toy} 
at the budgets in terms of different number of iterations. Given their different convergence behaviors,
when one wants to compromise with using pre-converged primal solutions in WLMs, 
he or she has to account for
the different results computed by different numerical methods, even though they all aim at the Wasserstein loss. 

As a note, Sinkhorn, B-ADMM and Gibbs-OT share the same computational complexity per iteration. 
The difference in their actual CPU time comes from the different arithmetic operations used. 
B-ADMM may be the slowest because it requires \texttt{log()} and \texttt{exp()} operations. 
When memory efficiency is of concern, both the implementations of Sinkhorn and Gibbs-OT 
can be modified to take only $O(m_1+m_2)$ additional memory besides the space for caching the cost matrix $M$.

\noindent\textbf{Two Electrons with Coulomb Cost in DFT.}
In quantum mechanics, Coulomb cost (or electron-electron Coulomb repulsion) is an important energy functional
in Density Functional Theory (DFT). Numerical methods that solve the multi-marginal OT problem 
with unbounded costs remains an open challenge in DFT~\cite{benamou2016numerical}.
We consider two uniform densities on 1D domain [0, 1] with Coulomb cost $c(x,y)=1/|x-y|$
which has analytic solutions. Coulumb cost is different from the usual metric cost in the 
OT literature, which is unbounded and singular at $x=y$. 
As observed in~\cite{benamou2016numerical}, the entropic regularized primal solution becomes more concentrated at 
boundaries, which is not physically plausible. This effect is not observed in the Gibbs-OT solution as shown in Appendix Fig.~\ref{fig:coulomb}. 
As shown by Fig~\ref{fig:sa}, the variables $\mathbf U, \mathbf V$ in computation are always
in bounded range (with an overwhelming probability), thus the algorithm does not endure any numerical difficulties. 

For entropic regularization~\cite{benamou2015iterative,benamou2016numerical}, 
we empirically select the minimal $\varepsilon$ which does not 
cause numerical overflow before 5000 iterations (in which $\varepsilon=0.5/N$). For Gibbs-OT, we use a geometric 
temperature scheme such that $T = 2.0 (1/l^4)^{n/l} / N$ at the $n$-th iteration, where $l$
is the max iteration number. For the unbounded Coulomb cost, Bregman ADMM~\cite{wang2014bregman} 
does not converge to a solution close to the true optimum. 

\subsection{Wasserstein NMF}
We now illustrate how the proposed Gibbs-OT can be used as a ready-to-plugin
inexact oracle for a typical WLM --- Wasserstein NMF~\cite{sandler2009nonnegative,roletfast}. The data parallelization 
of this framework is natural because the Gibbs-OT samplers subject to different instances are independent.

\noindent\textbf{Problem Formulation.}
Given a set of discrete probability measures $\{\Phi_i\}_{i=1}^n$ (data) over $\mathbb R^d$, we want to
estimate a model $\Theta=\{\Psi_k\}_{k=1}^K$,
such that for each $\Phi_i$, there exists a membership vector $\beta^{(i)}\in \Delta_K$:
$\Phi_i \approx \sum_{k=1}^K \beta_k^{(i)}\Psi_k$,
where each $\Psi_k$ is again a discrete probability measure to be estimated. 
Therefore, Wasserstein NMF reads
$\min_{\Theta, \Xi} \sum_{i=1}^n W \left(\Phi_i, \sum_{k=1}^K \beta_k^{(i)}\Psi_k\right)$,
where $\Xi=(\beta^{(1)},\ldots,\beta^{(n)})$ is the collection of membership vectors,
and $W$ is the Wasserstein distance. 
One can write the problem by plugging Eq.~\eqref{eq:dual} in the dual formulation:
\begin{eqnarray}
\min_{\Theta, \Xi} \max_{F=\{\mathbf f_i \}_{i=1}^n} && \!\!\!\!\! \sum\nolimits_{i=1}^n 
\left[\langle \widehat{\mathbf w}^{(i)},\mathbf g_i\rangle - \langle \mathbf w^{(i)},\mathbf h_i\rangle \right]\\
\mbox{s.t.} && \!\!\!\!\! \Psi_k = \sum\nolimits_{i=1}^m v_i^{(k)} \delta_{\mathbf x_i}\;,\\
&& \!\!\!\!\! \widehat{\Phi}^{(i)} = \sum\nolimits_{k=1}^K \beta_k^{(i)} \Psi_k\;,\\
&& \!\!\!\!\! \mathbf f_i \in \Omega\left(M(\widehat{\Phi}^{(i)}, \Phi_i)\right)\;,
\end{eqnarray}
where $\widehat{\mathbf w}^{(i)}\in \Delta_{m}$ 
is the weight vector of discrete probability measure $\widehat{\Phi}^{(i)}$
and $\mathbf w^{(i)}\in \Delta_{m_i}$ is the weight vector of $\Phi^{(i)}$.
$M(\cdot,\cdot)$ denotes the transportation cost matrix between the supports of two measures. 
The global optimization solves all three sets of variables $(\Theta, \Xi, F)$. 
In the sequel, we assume support points of $\{\Psi_k\}_{k=1}^m$ --- $\{\mathbf x_i\}_{i=1}^m$ are shared and pre-fixed.

\noindent\textbf{Algorithm.} At every epoch, one updates variables either sequentially (indexed by $i$) 
or all together. It is done by first executing the Gibbs-OT oracle subject to the $i$-th instance and 
then updating $\mathbf v^{(k)}$ and the membership vector $\beta^{(i)}$ accordingly at
a chosen step size $\gamma>0$. At the end of each epoch, the temperature parameter $T$ is adjusted 
$T := T\left(1-\sqrt{\frac{1}{m+\bar{m}}}\right)$,
where $\bar{m}=\frac{1}{n}\sum_{i=1}^n m_i$~. 
For each instance $i$, the algorithm proceeds with the following steps iteratively:
\begin{enumerate}
\item Initiate from the last computed $\mathbf U/\mathbf V$ sample subject to instance $i$, 
execute the Gibbs-OT Gibbs sampler at constant temperature $T$
until a mixing criterion is met, and get $\mathbf U_i$.
\item For $k=1,\ldots,K$, update $\mathbf v^{(k)}\in \Delta_{m}$ based on gradient 
$\beta_k^{(i)} \mathbf U_i$ using the iterates of online mirror descent (MD)
subject to the step-size $\gamma$ ~\cite{beck2003mirror}. 
\item Also update the membership vector $\mathbf \beta^{(i)}\in \Delta_{K}$ based on gradient 
$(\langle \mathbf v^{(1)}, \mathbf U_i\rangle,\ldots,\langle \mathbf v^{(K)}, \mathbf U_i\rangle)^T$ using 
the iterates of accelerated mirror descent (AMD) with restarts
subject to the same step-size $\gamma$ ~\cite{krichene2015accelerated}.
\end{enumerate}

\begin{figure}[ht!]\centering
\includegraphics[width=.5\textwidth, trim=2cm 9cm 2cm 8.5cm, clip]{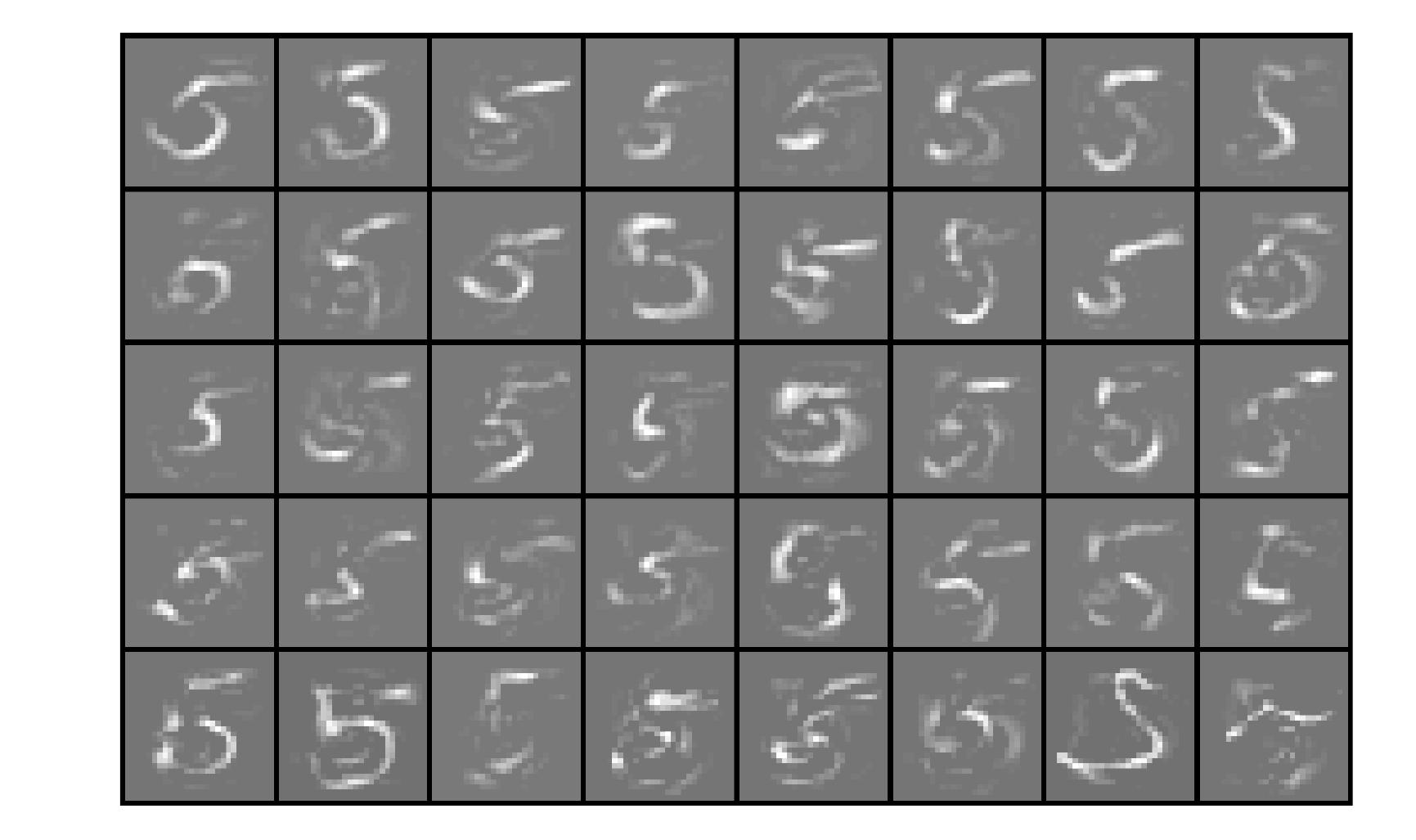}
\includegraphics[width=.5\textwidth, trim=2cm 9cm 2cm 8.5cm, clip]{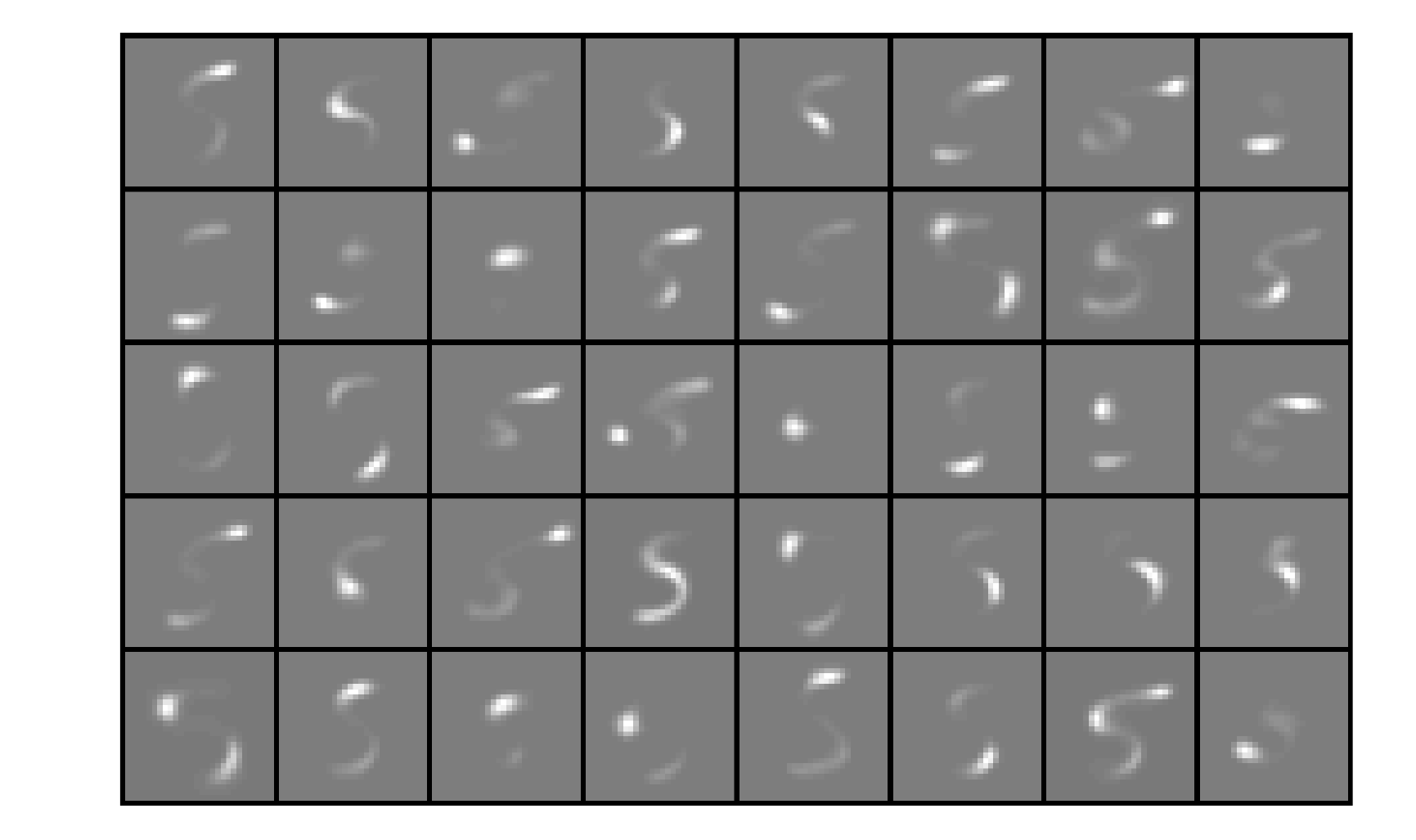}
\includegraphics[width=.5\textwidth, trim=2cm 9cm 2cm 8.5cm, clip]{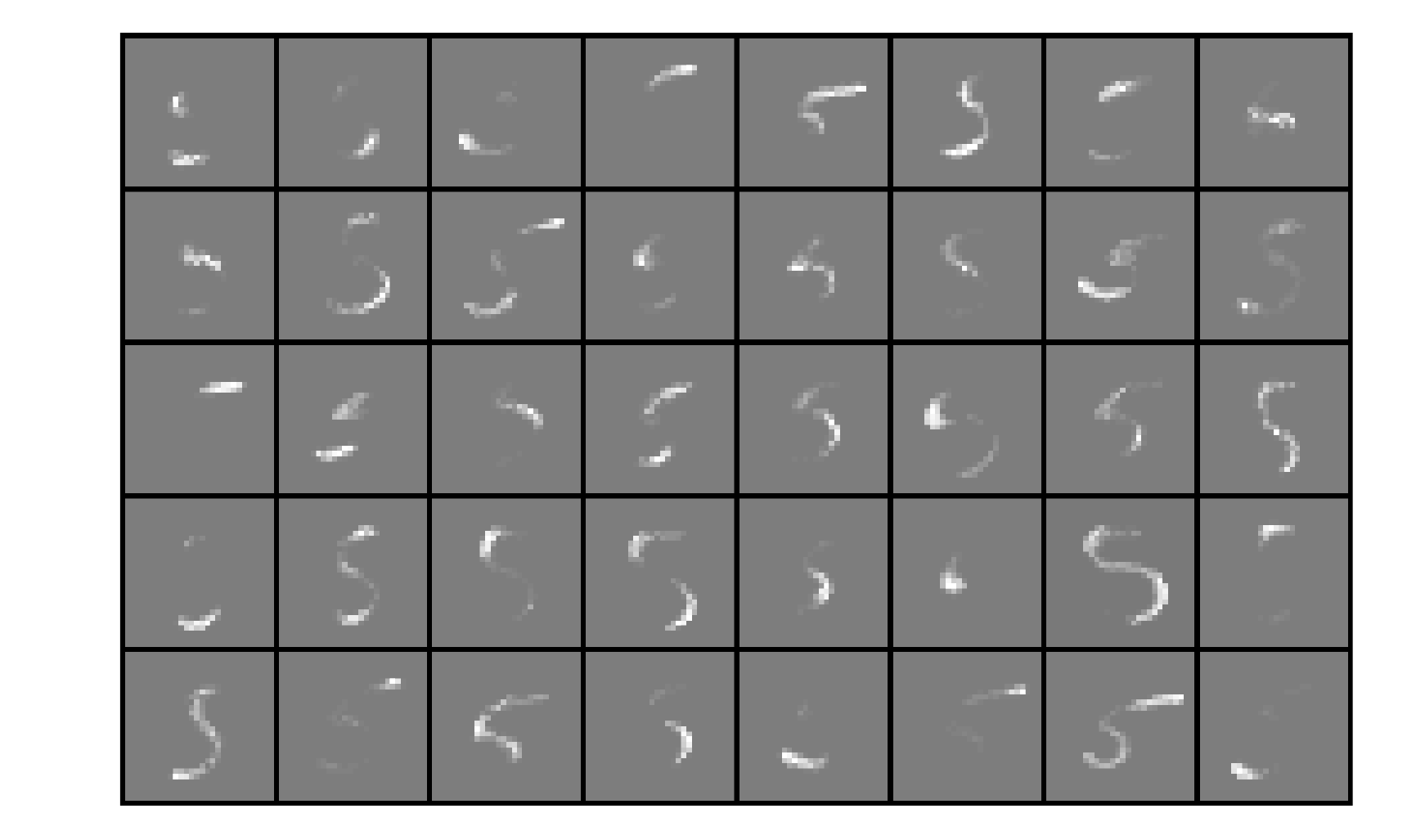}
\caption{NMF components learned by different methods ($K=40$) on the 200 digit ``5'' images. 
Top: regular NMF; Middle: W-NMF with entropic regularization ($\varepsilon=1/100$, $\rho_1=\rho_2=1/200$); 
Bottom: W-NMF using Gibbs-OT.
It is observed that the components of W-NMF with entropic regularization are
smoother than those optimized with Gibbs-OT.}\label{fig:mnist_nmf}
\end{figure}

\begin{figure}[ht!]\centering
\includegraphics[width=.5\textwidth, trim=2cm 8cm 2cm 8cm, clip]{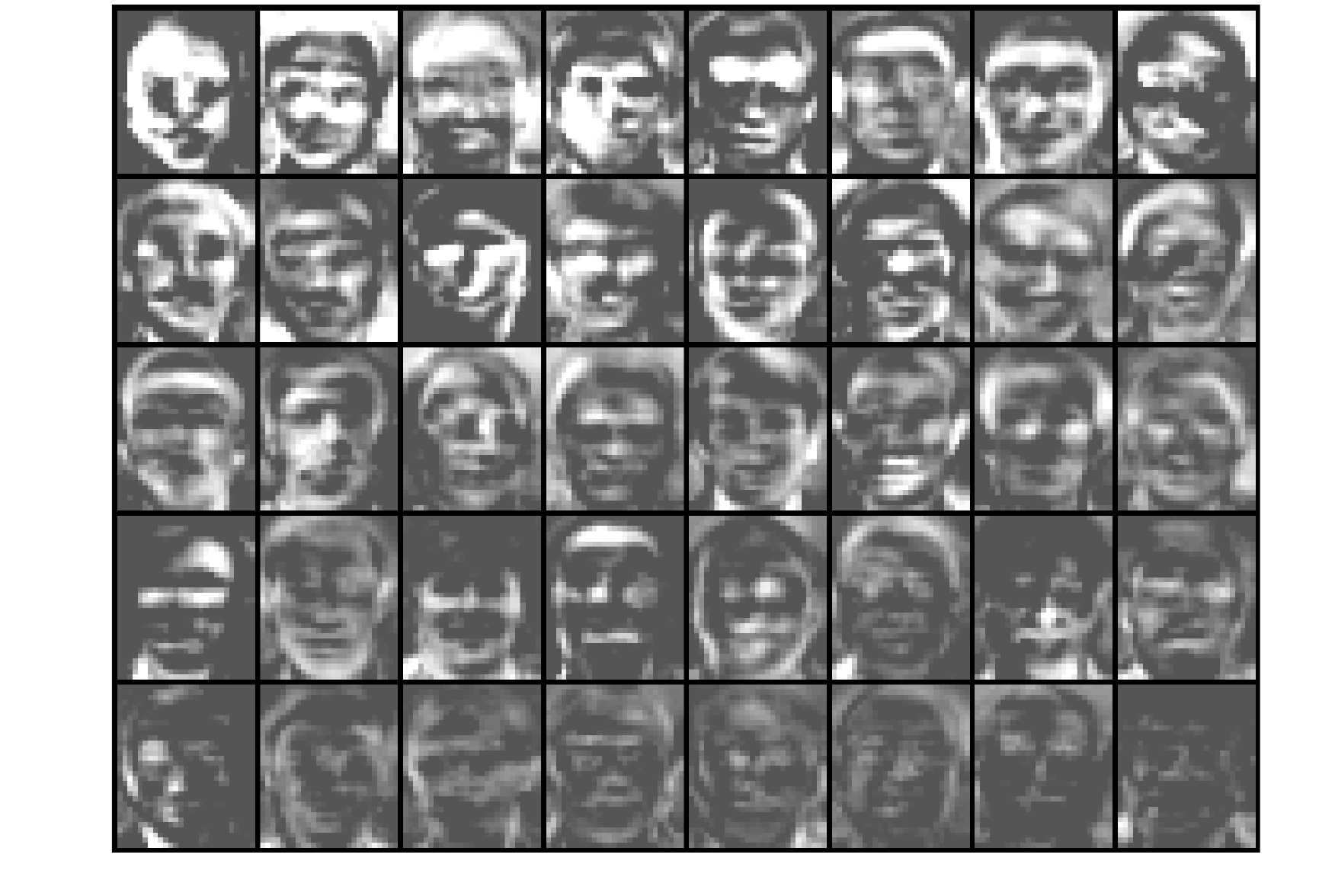}
\includegraphics[width=.5\textwidth, trim=2cm 8cm 2cm 8cm, clip]{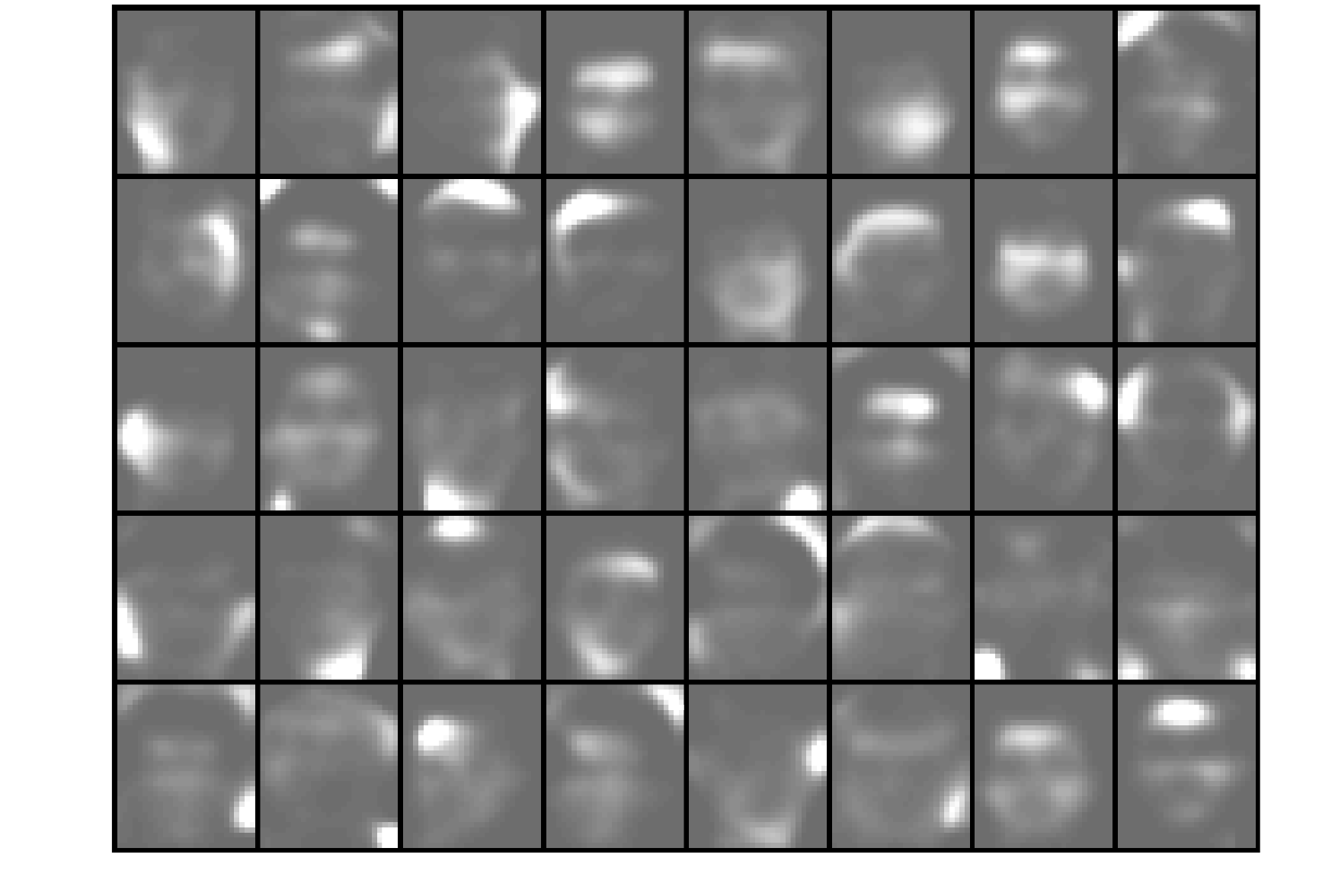}
\includegraphics[width=.5\textwidth, trim=2cm 8cm 2cm 8cm, clip]{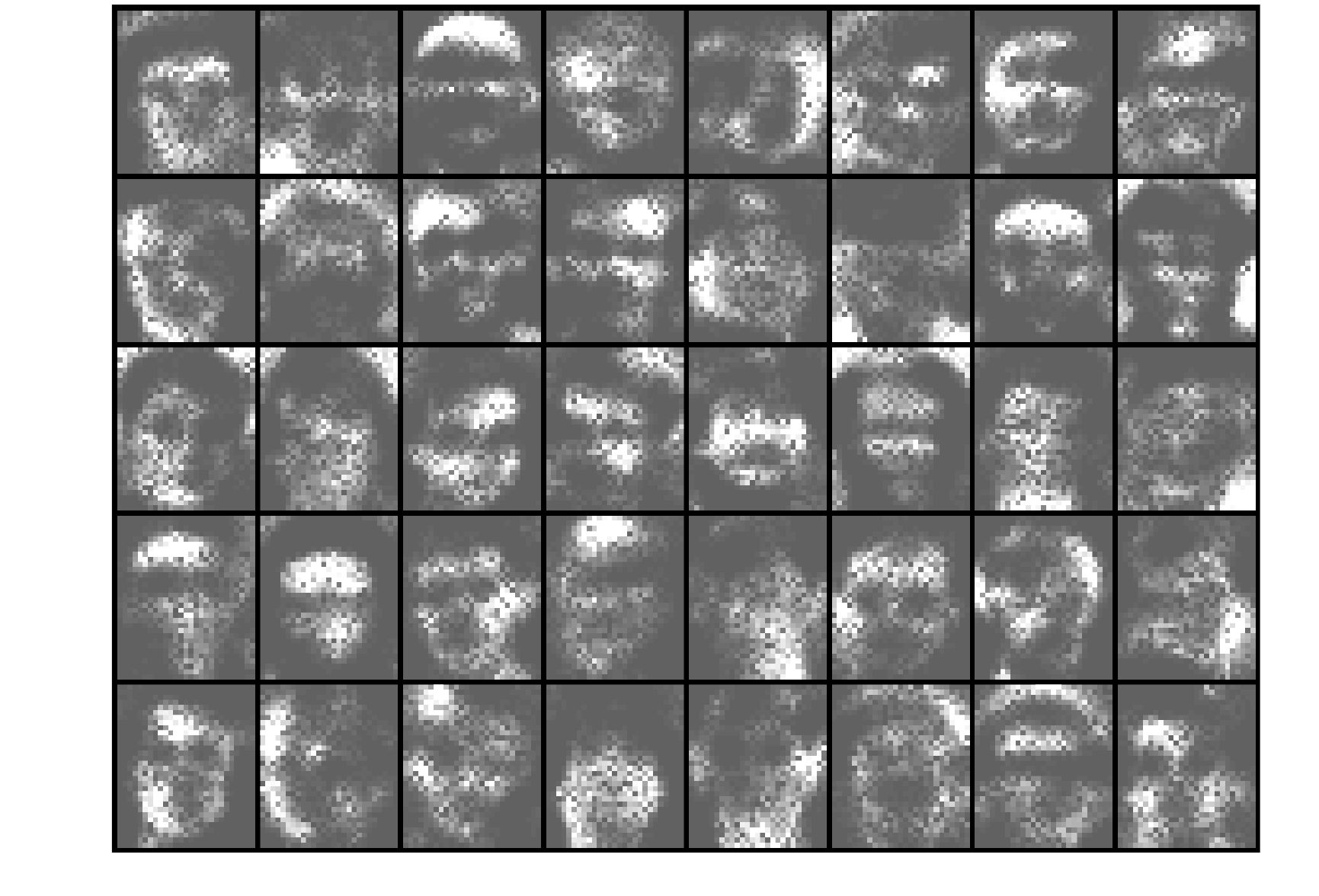}
\caption{NMF components learned by different methods ($K=40$) on the ORL face images. 
Top: regular NMF; Middle: W-NMF with entropic regularization ($\varepsilon=1/100$, $\rho_1=\rho_2=1/200$); 
Bottom: W-NMF using Gibbs-OT, in which the salt and pepper noises are observed due
to the fact that Wasserstein distance is insensitive to the subpixel 
mass displacement~\cite{cuturi2015smoothed}.}\label{fig:orl_nmf}
\end{figure}

We note that the practical speed-ups we achieved via the above procedure is the warm-start
feature in Step 1. If one uses a black-box OT solver, this dimension of speed-ups is not viable.

\noindent\textbf{Results.} We investigate the empirical convergence of the proposed Wasserstein NMF method by 
two datasets: one is a subset of MNIST handwritten digit images which contains 200 digits of ``5'',
and the other is the ORL 400-face dataset. Our results are based on a C/C++ implementation with vectorization. 
In particular, we set $K=40,\gamma=2.0$ for both datasets. 
The learned components are visualized
together with alternative approaches (smoothed W-NMF~\cite{roletfast} and regular NMF) 
in Appendix Figs.~\ref{fig:mnist_nmf} and~\ref{fig:orl_nmf}. From these figures, 
we observe that our learned components using Gibbs-OT are shaper than the smoothed W-NMF. 
This can be explained by the fact that Gibbs-OT can potentially
push for higher quality of approximation by gradually annealing the temperature.
We also observe that the learned components might possess some salt-and-pepper noise.
This is because the Wasserstein distance by definition is not very sensitive to the sub-pixel displacements. 
On a single-core of a 3.3 GHz Intel Core i5 CPU, 
the average time spent for each epoch for these two datasets are 0.84 seconds and 16.8 seconds, respectively.
It is about two magnitude faster than fully solving all OTs 
via a commercial LP solver~\footnote{We use the specialized network flow solver in Mosek (\url{https://www.mosek.com}) 
for the computation, which is found
faster than general simplex or IPM solver at moderate problem scale.}.

\section{Discussions}\label{sec:discuss}

The solution of primal OT (Monge-Kantorovich problem) have many direct interpretations, 
where the solved transport is a coupling between two measures. Hence, it could be well motivated
to consider regularizing the solution on the primal domain in those problems~\cite{cuturi2013sinkhorn}.
Meanwhile,
the solution of dual OT can be meaningful in its own right. 
For instance, in finance, the dual solution is directly interpreted as the vanilla prices 
implementing robust static super-hedging strategies. 
The entropy regularized OT, under the Fenchel-type dual, provides a 
smoothed unconstrained dual problem as shown in~\cite{cuturi2015smoothed}. 
In this paper, we develop Gibbs-OT,
whose solutions respect the dual feasibility of OT and are 
subject to a different regularization effect as explained by~\cite{abernethy2015faster}.
It is a numerical stable and computational suitable oracle to handle WLM. 

\textbf{Acknowledgement.}
This material is based upon work supported by the National Science Foundation under Grant Nos. ECCS-1462230 and DMS-1521092.
The authors would also like to thank anonymous reviewers for their valuable comments.

\bibliography{example_paper}
\bibliographystyle{icml2017}

\newpage
\normalsize
\appendix

\section{Theoretical Properties of Gibbs-OT}\label{sec:theory}
We develop quantitative concentration bounds for Gibbs-OT in a finite number of iterations
in order to understand the relationship between the temperature schedule and the concentration progress.
The analysis also guides us to adjust cooling schedule on-the-fly, as will be shown.  Proofs are provided in Supplement.

\noindent\textbf{Preliminaries.}
Before characterizing the properties of Gibbs-OT by Definition~\ref{alg:sa-ot}, we first give
the analytic expression for $p(\mathbf z^{n+1} | \mathbf z^{n})$. 
Let $G(\cdot): [-\infty, \infty] \mapsto [0,1]$ be the c.d.f. of standard exponential distribution.
Because $L_j^{(t+1)} < x$  by definition $\Leftrightarrow \forall i, \quad g_i^{(t)} - M_{i,j} < x$, 
the c.d.f. of $L_j^{(t+1)} | \mathbf U^{(t)}$ reads
\[\mbox{Pr}\left(L_j^{(t+1)}\!\! <\! x \left| \mathbf U^{(t)}\right.\right) \!=\! 
\prod_{i=1}^{m_1} \left( 1- G \!\left( \frac{- x - M_{i,j} + U_i^{(t)}}{ T^{(2t)} / p_i}\right)\right).\]
Likewise, the c.d.f. of $U_i^{(t)} | \mathbf L^{(t)}$ reads
\[\mbox{Pr}\left(U_i^{(t)} < x\left| \mathbf L^{(t)}\right.\right) \!=\! 
\prod_{j=1}^{m_2} G\!\left(\frac{x - M_{i,j} - L_j^{(t)}}{T^{(2t-1)}/q_j}\right).\]

With some calculation, the following can be shown. 
As a note, this lemma provides an intermediate result whose main purpose is to lay down the definition of $\phi_j^{(t)}$ and $\varphi_i^{(t)}$, which are then used in defining $O(z,T)$ (Eq.~\eqref{eq:ediff}) and $r^n$ (Eq.~\eqref{eq:r}) and in Theorem~\ref{bounds}.

\begin{lemma} \label{lemma0} (i)
Given $1\le j\le m_2$ and $1\le t\le N$, let the sorted index of $\{U_i^{(t)} - M_{i,j}\}_{i=1}^{m_1}$ be
permutation $\{\sigma(i)\}_{i=1}^{m_1}$ such that sequence $\{U_{\sigma(i)}^{(t)}-M_{\sigma(i),j}\}_{i=1}^{m_1}$ 
are monotonically non-increasing. Define the auxiliary quantity 
\begin{equation}
\phi_j^{(t)} \myeq \sum_{k=1}^{m_1}
\dfrac{(1-\mu_{k}) \prod_{i=1}^{k-1}\mu_i }{\sum_{i=1}^k p_{\sigma(k)}}\;,\label{eq:phi}
\end{equation}
where
\begin{multline*}
1 \ge \mu_i \myeq \exp \Bigg\{\dfrac{\sum_{i=1}^k p_{\sigma(k)}}{T^{(2t)}}\Big[ \left(U_{\sigma(i+1)} - M_{\sigma(i+1),j}\right)
\\ - \left(U_{\sigma(i)} - M_{\sigma(i),j}\right)\Big]\Bigg\}
\end{multline*} for $i=1,\ldots,m_1-1$, and $\mu_{m_1} \myeq 0$~.  
Then, the conditional expectation
\[\mathbb E \left[L_j^{(t+1)} \left| \mathbf U^{(t)}\right.\right] = U_{\sigma(1)}^{(t)} - M_{\sigma(1),j} - \phi_j^{(t)} T^{(2t)}\;.\]
In particular, we denote $\sigma(1)$ by $I_j^t$ or $I(j,t)$\;. 

(ii) Given $1\le i \le m_1$ and $1\le t \le N$, let the sorted index of $\{M_{i,j} + L_j\}_{j=1}^{m_2}$ 
be permutation $\{\sigma(j)\}_{j=1}^{m_2}$ such that the sequence $\{M_{i,\sigma(j)}+L_{\sigma(j)}^{(t)}\}_{j=1}^{m_2}$
are monotonically non-decreasing. Define the auxiliary quantity
\begin{equation}
\psi_i^{(t)} \myeq \sum_{k=1}^{m_2} \dfrac{(1-\lambda_k) \prod_{j=1}^{k-1} \lambda_k}{\sum_{j=1}^k q_{\sigma(j)}}\;,\label{eq:psi}
\end{equation}
where 
\begin{multline*}
1 \ge  \lambda_j\myeq \exp\Bigg\{\dfrac{\sum_{j=1}^k q_{\sigma(j)}}{T^{(2t-1)}}\Big[\left(M_{i,\sigma(j)}+L_{\sigma(j)}^{(t)}\right) \\ 
-\left(M_{i,\sigma(j+1)}+L_{\sigma(j)}^{(t+1)}\right)\Big]\Bigg\} 
\end{multline*} for $i=1,\ldots,m_2-1$ and $\lambda_{m_2}=0$~.
Then, the conditional expectation \[
\mathbb E \left[U_i^{(t)} \left| \mathbf L^{(t)}\right.\right] =M_{i,\sigma(1)}+L_{\sigma(1)}^{(t)} +\psi_i^{(t)} T^{(2t-1)}\;.\]
In particular, we denote $\sigma(1)$ by $J_i^t$ or $J(i,t)$~. 
\end{lemma}
We note that the calculation of Eq.~\eqref{eq:phi} and Eq.~\eqref{eq:psi} needs $O(m_1\log m_1)$ 
and $O(m_2\log m_2)$ time respectively. By a few additional calculations, 
we introduce the notation $\mathcal O(\cdot,\cdot)$: 
\begin{eqnarray}
&&\!\!\!\!\!\!\!\!\!\! \mathcal O(\mathbf z^{2t}, T^{(2t)}) \nonumber \\ 
&&\!\!\!\!\!\!\!\!\!\! \myeq\mathbb E \left[\langle\mathbf q, \mathbf L^{(t)}\rangle - \langle \mathbf q, \mathbf L^{(t+1)}\rangle 
\left| \mathbf U^{(t)}, \mathbf L^{(t)} \right.\right] 
\nonumber \\
&&\!\!\!\!\!\!\!\!\!\!=\sum_{j=1}^{m_2} \left(M_{I_j^t,j} + L_j^{(t)} - U_{I_j^t}^{(t)} + \phi_j^{(t)} T^{(2t)}\right) q_j  \nonumber\\
&&\!\!\!\!\!\!\!\!\!\!\mathcal O(\mathbf z^{2t-1}, T^{(2t-1)}) \nonumber \\
&&\!\!\!\!\! \!\!\!\!\!\myeq\mathbb E \left[\langle\mathbf p, \mathbf{U}^{(t)}\rangle - \langle\mathbf p, \mathbf{U}^{(t-1)}\rangle 
\left| \mathbf U^{(t-1)}, \mathbf L^{(t)} \right.\right] \nonumber\\
&&\!\!\!\!\! \!\!\!\!\!=\sum_{i=1}^{m_1} \left(M_{i,J_i^t} + L_{J_i^t}^{(t)} - U_{i}^{(t-1)} + \psi_i^{(t)} T^{(2t-1)}\right) p_i \nonumber\\ \label{eq:ediff}
\end{eqnarray}
Note that $\mathcal O(\mathbf z^n, T^n) = \mathbb E \left[V(\mathbf z^{n+1})-V(\mathbf z^n) | \mathbf z^n\right]$~.

\noindent\textbf{Recovery of Approximate Primal Solution.} 
An approximate $(m_1+m_2)$-sparse primal solution\footnote{The notation of $\mbox{sparse} (\cdot, \cdot, \cdot)$ function is introduced under the syntax of MATLAB: \url{http://www.mathworks.com/help/matlab/ref/sparse.html}}
can be recovered from $\mathbf z^n$ at $n=2t$ by
\begin{multline}
Z \approx \dfrac{1}{2} \mbox{sparse}(1:m_1, J(1:m_1,t), \mathbf p) +\\
\dfrac{1}{2} \mbox{sparse}(I(1:m_2,t), 1:m_2, \mathbf q) \in \mathbb R^{m_1\times m_2}.
\end{multline}

\noindent\textbf{Concentration Bounds.}
We are interested in the concentration bound
related to $ V(\mathbf z^{n}) $ because it replaces the true Wasserstein loss in WLMs. 
Given $\mathbf U^{(0)}$ (\textit{i.e.}, $\mathbf z^1$ is implied), 
for $n=1,\ldots,2N$, we let 
\begin{equation}
r^{n} = V(\mathbf z^n) 
- \sum_{s=1}^{n-1} \mathcal O(\mathbf z^{s}, T^{(s)})\;.\label{eq:r}
\end{equation}
This is crucial for one who wants to know whether the cooling schedule is too
fast to secure the suboptimality within a finite budget of iterations. 
The following Theorem~\ref{bounds} gives a possible route to approximately realize this goal. 
It bounds the difference between 
\[V(\mathbf z^n) - V(\mathbf z^1) \mbox{ and } \sum_{s=1}^{n-1} \mathbb E \left[V(\mathbf z^{s+1})-V(\mathbf z^s)| \mathbf z^s\right],\]
the second of which is a quantitative term representing sum of a sequence. 
We see that $\mathcal O(\mathbf z^{s}, T^{(s)}) = 
\mathbb E \left[V(\mathbf z^{s+1})-V(\mathbf z^s)| \mathbf z^s\right] =0$ if and only if $T^{(s)} = \mathcal T (\mathbf z^s) \myeq $
\begin{equation}
\begin{cases}
-\dfrac{1}{\langle \phi^{(t)}, \mathbf q\rangle}  \sum\limits_{j=1}^{m_2}q_j\left[ M_{I_j^t,j} + L_j^{(t)} - U_{I_j^t}^{(t)} \right] &\mbox{if } s\!=\!2t\\
-\dfrac{1}{\langle \psi^{(t)}, \mathbf p\rangle}  \sum\limits_{i=1}^{m_1}p_i\left[ M_{i,J_i^t} + L_{J_i^t}^{(t)} - U_{i}^{(t-1)}\right] & \mbox{if } s\!=\!2t\!-\!1
\end{cases}\label{eq:tbound}
\end{equation}
In the practice of Gibbs-OT, choosing the proper cooling schedule for a specific WLM needs trial-and-error. Here
we present a heuristics that the temperature $T^{(s)}$ 
is often chosen and adapted around $\eta \mathcal T(z^s)$, where $\eta\in [0.1, 0.9]$. 
We have two concerns regarding the choice of temperature $T$: 
First, in a WLM, the cost $V(\mathbf z)$ is to be gradually minimized,
hence a temperature $T$ smaller than $\mathcal T(\mathbf z^{s})$ at every iteration 
ensures that the cost is actually decreased by expectation, 
\textit{i.e.}, $\mathbb E[V(\mathbf z^n) - V(\mathbf z^{1})] < 0$; 
second, if $T$ is too small, it takes many iterations to reach
a highly accurate equilibrium, which might not be necessary for a single outer level step of parameter update.

\begin{theorem}[Concentration bounds for finite time Gibbs-OT]\label{bounds}
First, $r^{n}$ (by definition) is a martingale subject to the filtration of $\mathbf z_1, \ldots, \mathbf z_n$. 
Second, given a $\varepsilon\in (0,1)$, for $n=1,\ldots,2N-1$
if we choose the temperature schedule $T^{(1)},\ldots,T^{(2N)}$ such that 
(i) $C^n \cdot T^{(n)} \le a_n$, or
(ii) $\exists \gamma>0$, $ \log\left(\frac{ 2N\max\{m_1,m_2\}}{\varepsilon}\right) \cdot T^{(n)}+ D^{n} \le \gamma a_n$,
where $\{a_n \ge 0\}$ is a pre-determined array.
Here for $t=1,\ldots,N,$
\begin{eqnarray*}
C^{2t-1} &\myeq& \langle \psi^{(t)}, \mathbf p\rangle\;, \\
C^{2t} &\myeq& \langle \phi^{(t)}, \mathbf q\rangle\;,\\
D^{2t-1} &\myeq& \sum_{i=1}^{m_1} p_i\mathcal R\left(M_{i,\cdot}^T + \mathbf L^{(t)};\mathbf q\right)\;,\\
D^{2t} &\myeq& \sum_{i=1}^{m_2} q_j \mathcal R\left(M_{\cdot,j} - \mathbf U^{(t)}; \mathbf p\right)\;,
\end{eqnarray*}
where $M_{i,\cdot}$ and $M_{\cdot,j}$ represents the $i$-th rows and $j$-th columns of matrix $M$ respectively,
$\psi^{(t)}$ and $\phi^{(t)}$ are defined in Lemma~\ref{lemma0},
and regret function $\mathcal R(\mathbf x; \mathbf w) \myeq  \sum_{i=1}^{m} w_i x_i - \min_{1\le i\le m} x_i$
for any $\mathbf w\in \Delta_m$ and $\mathbf x\in \mathbb R^m$. 
Then for any $K>0$, we have
\begin{eqnarray}
&&\!\!\!\!\!\!\!\!\!\!\!\!\!\!\!\mbox{Pr}\left(r^{2N} \!<\! r^{1} \!-\!K \right) \le \exp\left[-\dfrac{K^2}{2\sum_{i=1}^{2N-1} a_n^2}\right]\;,
\label{eq:leftbound}\\
&&\!\!\!\!\!\!\!\!\!\!\!\!\!\!\!\mbox{or}\nonumber\\
&&\!\!\!\!\!\!\!\!\!\!\!\!\!\!\!\mbox{Pr}\left(r^{2N}   \!>\! r^{1} \!+\!\gamma K \right) \le \exp\left[-\dfrac{K^2}{2 \sum_{i=1}^{2N-1} a_n^2} \right]\!\!+ \varepsilon\;.\label{eq:rightbound}
\end{eqnarray}
\end{theorem}
\begin{remark}
The bound obtained is a quantitative Hoeffding bound,
not a bound that guarantees contraction around the true solution of dual OT. 
Nevertheless, we argue that this bound is still useful
in investigating the proposed Gibbs sampler when the temperature is not annealed to zero. 
Particularly, the bound is for cooling schedules in general, {\it i.e.}, it is more applicable than a bound for a specific schedule.
There has long been a gap between the practice and theory of SA despite of its wide usage. 
Our result likewise falls short of firm theoretical guarantee from the optimization perspective, 
as with the usual application of SA.
\end{remark}

\section{Proof of Lemmas and Theorem}
The minimum of $n$ independent exponential random variables with different parameters 
has computatable formula for its expectation. The result immediately lays out the proof of 
Lemma~\ref{lemma0}. 
\begin{lemma}\label{lemma_a0}
Suppose we have $n$ independent exponential random variables $e_i$ whose c.d.f.
is by $f_i(x) = \min\{
\exp (\omega_i (x-z_i)),1\}$. Without lose of generality, we assume $z_1\ge z_2 \ge ... \ge z_n$, then let $z_{n+1}=-\infty,
h_i=\exp\left[\sum_{j=1}^i \omega_j (z_{i+1}-z_i)\right] \le 1$ (with $h_n=0,z_{n+1}h_{n}=0$), we have
\[\mathbb E\left[\max \{e_1,\ldots,e_n\}\right] =  z_1 - \sum_{i=1}^n \dfrac{(1-h_i)\prod_{j=1}^{i-1}h_i}{\sum_{j=1}^i \omega_j}\;.\]
\begin{proof}
The c.d.f. of $\max \{e_1,\ldots,e_n\}$ is 
$F(x)=\prod_{i=1}^n f_i(x)$ which is piece-wise smooth with interval $(z_{i+1}, z_i)$, 
we want to calculate $\int_{-\infty}^{\infty} x dF(x)$~.
\begin{eqnarray*}
&&\int_{-\infty}^{\infty} x dF(x) \\
&=&\sum_{i=1}^{n} \int_{z_{i+1}}^{z_i} x dF(x) + 0\\
&=& \sum_{i=1}^{n} \int_{z_{i+1}}^{z_i} x d \exp\left[\sum_{j=1}^{i}\omega_j (x-z_j) \right]\\
&=& \sum_{i=1}^{n} \int_{z_{i+1}}^{z_i} \left[\sum_{j=1}^i \omega_j \right] x \exp\left[\sum_{j=1}^{i}\omega_j (x-z_j)  \right] dx\\  
&=& \sum_{i=1}^n \Big\{\left(z_i - \dfrac{1}{\sum_{j=1}^i \omega_j } \right)\exp\left[\sum_{j=1}^{i}\omega_j (z_i-z_j) \right] \\
&& - \left(z_{i+1} - \dfrac{1}{\sum_{j=1}^i\omega_j } \right)\exp\left[\sum_{j=1}^{i}\omega_j (z_{i+1}-z_j) \right]\Big\}\\
&=& \sum_{i=1}^n  
\left[\left( z_i - z_{i+1} h_i \right) - \dfrac{1-h_i}{\sum_{j=1}^i \omega_j }\right]\prod_{j=1}^{i-1} h_i\\
&=& \sum_{i=1}^n \left[z_i \prod_{j=1}^{i-1} h_i - z_{i+1}\prod_{j=1}^{i} h_i\right]\\
&&-\sum_{i=1}^n \dfrac{(1-h_i)\prod_{j=1}^{i-1}h_i}{\sum_{j=1}^i \omega_j} \\
&=& z_1 - \sum_{i=1}^n \dfrac{(1-h_i)\prod_{j=1}^{i-1}h_i}{\sum_{j=1}^i \omega_j}\;.
\end{eqnarray*}
\end{proof}
\end{lemma}
Therefore Lemma~\ref{lemma0} is proved up to trivial calculation using the above Lemma~\ref{lemma_a0}. 
In order to further prove Lemma~\ref{lemma1}, we also have (by definition of $F(x)$).
\begin{lemma}\label{lemma_a1}
Subject to the setup of Lemma~\ref{lemma_a0}, we also have 
\[\max \{e_1,\ldots,e_n\} \le z_1\;,\]
and 
\[F(x) \le \min\left\{\exp\left[\sum_{i=1}^n\omega_i (x - z^\ast)\right], 1\right\}, -\infty < x < \infty,\]
where $z^\ast = \dfrac{\sum_{i=1}^n \omega_i z_i }{\sum_{i=1}^n \omega_i}$~. 
\end{lemma}
Therefore, based on the observation of Lemma~\ref{lemma_a1}, the tail probability 
$Pr(\max \{e_1,\ldots,e_n\} < x)$ is upper bounded by the probability 
of an exponential random variable, which lead us to the proof of Lemma~\ref{lemma1}.
\begin{lemma} \label{lemma1}
Note that Eq.~\eqref{eq:ediff} implies $\mathbb E\left[r^{n+1}-r^{n} | \mathbf z^{1}, \ldots, \mathbf z^n\right] = 0$
for $t=1,\ldots,2N$. Therefore, $\{r^n\}$ is a (discrete time) martingale subject to the filtration of $\{\mathbf z^n\}$.
(Recall the notation by Eq.~\eqref{eq:z}.) Moreover, we have the following two bounds.
First, we can establish the left hand side bound for $\{r^{n+1} - r^n\}_{n=1}^{2N-1}$:
\[r^{n}-r^{n+1} \le C^n \cdot T^{(n)}, \]
where for $t=1,\ldots,N$
\begin{equation}
C^{2t-1} \myeq \langle \psi^{(t)}, \mathbf p\rangle \mbox{ and }
C^{2t}\myeq \langle \phi^{(t)}, \mathbf q\rangle.
\end{equation}
Second, we also bound on the right hand side. That said,
for any $1>\varepsilon>0$, we have
\begin{multline}
\mbox{Pr}\Big(\exists n\in \{1,\ldots,2N\},\mbox{ s.t. } r^{n+1} - r^{n} \\ \ge \log\left(\frac{ 2N \max\{m_1,m_2\}}{\varepsilon}\right) \cdot T^{(n)}+ D^{n} \Big|  \mathbf z^{1}, \ldots, \mathbf z^n \Big) \le\varepsilon,
\end{multline}
where for $t=1,\ldots,N$
\begin{eqnarray}
D^{2t-1} \myeq \sum_{i=1}^{m_1} p_i\mathcal R\left(M_{i,\cdot}^T + \mathbf L^{(t)};\mathbf q\right)
\\ 
D^{2t} \myeq \sum_{i=1}^{m_2} q_j \mathcal R\left(M_{\cdot,j} - \mathbf U^{(t)}; \mathbf p\right),
\end{eqnarray}
where $M_{i,\cdot}$ and $M_{\cdot,j}$ represents the $i$-th rows and $j$-th columns of matrix $M$ respectively. 
\begin{proof}
On one hand, because for each $i\in\{1,\ldots,m_1\}$, $ U_i^{(t)} | \mathbf L^{(t)}$ 
is lower bounded by $M_{i, J(i,t)} + L_{J(i,t)}^{(t)} $ (Lemma~\ref{lemma_a1}),
and for each $j\in\{1,\ldots,m_2\}$, 
$L_j^{(t)} | \mathbf U^{(t-1)}$ is upper bounded by $U_{I(j,t)}^{(t-1)}-M_{I(j,t),j}$ (Lemma~\ref{lemma_a1}),
we easily (by definition) have $r^{n+1} | \mathbf z_1, \ldots,\mathbf z^n$ is lower bounded
by $r^n - C^n \cdot T^{(n)}$. 

On the other hand, we have
if $ r^{n+1}-r^n \ge \log(1/\varepsilon_0)\cdot T^{(n)} +D^n | \mathbf z_1,\ldots,\mathbf z_n$ for some $\varepsilon_0>0$, then at least 
one of $U_i^{(t)}$ (or $L_j^{(t)}$) violates the bound 
$\log(1/\varepsilon_0)\cdot T^{(n)} + \mathcal R(M_{i,\cdot}^T + \mathbf L^{(t)};\mathbf q)$
(or $\log(1/\varepsilon_0)\cdot T^{(n)} + \mathcal R(M_{\cdot,j}-\mathbf U^{(t)};\mathbf p)$),
whose probability using Lemma~\ref{lemma_a1} is shown to be less than $\varepsilon_0$. Therefore, we have for each $n$
\begin{multline}
Pr(r^{n+1}-r^n \ge \log(1/\varepsilon_0)\cdot T^{(n)} +D^n | \mathbf z_1,\ldots,\mathbf z_n)\\
\le \max\{m_1,m_2\} \varepsilon_0\;, \end{multline}
and 
\begin{multline}Pr(\exists n, r^{n+1}-r^n \ge \log(1/\varepsilon_0)\cdot T^{(n)} +D^n | \mathbf z_1,\ldots,\mathbf z_n)\\
\le 2N \max\{m_1,m_2\} \varepsilon_0\;,\end{multline}
Let $\varepsilon= 2N \max\{m_1,m_2\} \varepsilon_0\;$, which concludes our result. 
\end{proof}
\end{lemma}

Given Lemma~\ref{lemma1}, we can prove Theorem~\ref{bounds} by applying the classical Azuma's inequality
for the left-hand side bound, and 
applying one of its extensions (Proposition 34 in (Tao and Vu, 2015)) for the right-hand side bound.
Remark that Theorem~\ref{bounds} is about a single OT. For multiple different OTs,
which share the same temperature schedule, one can have asymptotic bounds using the Law of Large Numbers
due to the fact that their Gibbs samplers are independent with each other.
Let $R^n=\frac{1}{S}\sum_{k=1}^S {r_k^n}$~, where $r_k^n$ is defined
by Eq.~\eqref{eq:r} for sample $k$. Since for any $\varepsilon>0$, one has 
$P(\left|R^{n+1}-R^n\right|>\varepsilon) \to 0$~, 
as $S\to \infty$, one can have the asymptotic concentration bound for $R^{2N}$ that for 
any $\varepsilon_1,\varepsilon_2>0$~,
there exists $S$ such that
$P(\left|R^{2N} - R^1\right|>\varepsilon_1)\le \exp\left(-\frac{1}{2N \varepsilon_2}\right)$~. 

\vskip 0.2in
{\small Tao, Terence and Vu, Van. Random matrices: Universality of local spectral statistics of non-Hermitian matrices. {\it The Annals of Probability}, 43(2):782-874, 2015.}
\end{document}